\documentclass{article}
\usepackage{epsfig}
\usepackage{graphicx}
\def\d{\mbox{\rm d}}
\def\NN{\mbox{\tiny\rm NN}}
\def\be{\begin{equation}}
\def\ee{\end{equation}}
\def\bea{\begin{eqnarray}}
\def\eea{\end{eqnarray}}
\def\ave#1{\langle #1 \rangle}
\def\dst{\displaystyle\strut}

\def\dst{\displaystyle{\phantom{|}}}
\def\ov{\over\dst}

\begin{document}
\title{{\bf  A hint at quark deconfinement \\
	in 200 GeV Au+Au data at RHIC} }
\author{
M. Csan\'ad$^1$, T. Cs\"org\H{o}$^2$, B. L\"orstad$^3$ and A. Ster$^2$\\[2.0ex]
$^1$Dept. Atomic Phys., ELTE, H-1117 Budapest, P\'azm\'any P. 1/a, Hungary \\
$^2$MTA KFKI RMKI, H - 1525 Budapest 114, P.O.Box 49, Hungary \\
$^3$Dept. Physics, University of Lund, S - 22362 Lund, Sweden
}
\maketitle

\begin{abstract}
We give the emission function of the axially symmetric Buda-Lund
hydro model and present its simultaneous, high quality
fits to identified particle spectra, two-particle Bose-Einstein or
HBT correlations and charged particle pseudorapidity
distributions as measured by BRAHMS  and PHENIX  
in 0-30 \% central, $\sqrt{s_{\NN}} = 200$ GeV Au+Au collisions at RHIC.
The best fit is achieved when the most central region  
of the particle emitting volume is superheated to  $T_0 = 200 \pm 9$ MeV 
$ \ge  T_c =172 \pm 3$ MeV, a preliminary, 3 $\sigma$ effect. 
\end{abstract}

\section{Introduction}
The Buda-Lund hydro model is successful in describing the 
identified single particle spectra and the transverse mass dependent
Bose-Einstein or HBT radii as well as the pseudorapidity distribution
of charged particles in Au + Au collisions 
at $\sqrt{s_{\NN}} = 130 $ GeV~\cite{ster-ismd03},
as measured by the BRAHMS, PHENIX, PHOBOS and STAR collaborations.
The result of the simultaneous fit to all these datasets
indicate the existence of a very hot region, with
a temperature significantly greater than 170 MeV~\cite{mate-ell1}.
Recently, Fodor and Katz calculated the phase diagram
of lattice QCD  at finite net barion density~\cite{Fodor:2001pe}.
These lattice results, obtained with light quark masses 
four times heavier than the physical value,
indicated that in the $0 \le \mu_B \le 700$ MeV region the transition
from confined to deconfined matter is a cross-over,
with $T_c \simeq 172 \pm 3$ MeV. This value is, 
within one standard deviation, independent of the bariochemical 
potential in the $0 \le \mu_B \le 300$ MeV region. 
The Buda-Lund fits, combined with these lattice results,
provide an indication  for quark deconfinement 
in  Au + Au collisions with $\sqrt{s_{\NN}} = 130 $ GeV 
colliding energies at RHIC. This observation was
confirmed~\cite{mate-ell1} by the analysis of 
the transverse momentum and rapidity dependence of the elliptic flow
as measured by the PHENIX and PHOBOS collaborations.

Here we investigate what happens if 
a similar analysis is performed on the final, published
Au+Au collision data at RHIC at the maximum, $\sqrt{s_{\NN}} = 200$ GeV 
bombarding energies.

\section{The emission function of the Buda-Lund hydro model}
The Buda-Lund hydro model was introduced
in refs.~\cite{Csorgo:1995bi,Csorgo:1995vf}.
This model was defined in terms of its emission
function $S(x,k)$, for axial symmetry, corresponding to 
central collisions of symmetric nuclei. 
The observables are calculated analytically, 
see refs.~\cite{cs-rev,ster-ismd03} for details and key features.
Here we summarize the Buda-Lund emission function 
 in terms of its fit parameters. 
The presented form is equivalent
to the original shape proposed in refs.~\cite{Csorgo:1995bi,Csorgo:1995vf},
however, it is easier to fit and interpret it.

The single particle invariant momentum distribution, $N_1(k_1)$,
is obtained as 
\be
	N_1(k_1) = \int \d^4 x \, S(x,k_1).
\ee
For chaotic (thermalized) sources, in case of the validity of
the plane-wave approximation, the two-particle  
invariant momentum distribution $N_2(k_1,k_2) $
is also determined  by $S(x,k)$,
the single particle emission function, if non-Bose-Einstein
correlations play negligible role or can be corrected for,
see ref.~\cite{cs-rev} for a more detailed discussion. 
Then the two-particle Bose-Einstein correlation function,
$
C_2 (k_1,k_2) = { N_2(k_1,k_2)}/\left[{ N_1(k_1) N_1(k_2) }\right]
$
can be evaluated in a core-halo picture~\cite{Csorgo:1994in}, 
where the emission function
is a sum of emission functions characterizing a hydrodynamically evolving
core and a surrounding halo of decay products of long-lived resonances,
$S(x,k) = S_c(x,k) + S_h(x,k)$. Consequently, the single particle spectra
can also be given as a sum, $N_1(k) = N_{1,c}(k) + N_{1,h}(k)$. 
In the correlation function, an 
effective intercept parameter $\lambda \equiv \lambda_*(K)$ 
appears and its relative momentum dependence
can be calculated directly from the emission function of the core,
\be
C_2 (k_1,k_2) = 1+\frac{|\tilde S(q,K)|^2}
{|\tilde S(0,K)|^2} \simeq
1+\lambda_{*}(K)\frac{|\tilde S_c(q,K)|^2}{|\tilde S_c(0,K)|^2},
\ee
where the relative and the  momenta are 
$q  =  k_1-k_2$, 
$K  =   0.5 (k_1+k_2)$,
and the Fourier-transformed emission function is defined as
$\tilde S(q,K)  = \int \d^4 x S(x,K) \exp(i q x).$

The measured  $\lambda_*$ parameter of the
correlation function is utilized to correct the core spectrum for 
long-lived resonance decays~\cite{Csorgo:1994in}:
$ N_1(k) = N_c(k)/{\sqrt{\lambda_{*}(k)}}.  $
The emission function of the core is assumed to have a hydrodynamical form,
\begin{equation}
  S_c(x,k) d^4 x = \frac{g}{(2 \pi)^3}
  \frac{ k^\nu d^4\Sigma_\nu(x)}{B(x,k) +s_q},
\end{equation}
where $g$ is the degeneracy factor ($g = 1$ for 
pseudoscalar mesons, $g = 2$ for spin=1/2 barions).
The particle flux over the freeze-out layers is given by a generalized
Cooper--Frye factor: the freeze-out hypersurface depends
parametrically on the freeze-out time $\tau$ and the probability
to freeze-out at a certain value is proportional to $H(\tau)$,
$
         k^\nu d^4\Sigma_\nu(x)  = 
         m_t \cosh(\eta - y)  
        H(\tau) d\tau \, \tau_0 d\eta \, dr_x \, dr_y.
$
Here $\eta = 0.5 \log[(t + r_z)/(t-r_z)]$, $\tau=\sqrt{t^2 - r_z^2}$,
$ y = 0.5 \log[(E + k_z)/(E-k_z)]$ and $m_t=\sqrt{E^2 - k_z^2}$. 
The freeze-out time distribution $H(\tau)$ is approximated by a Gaussian,
$
        H(\tau) = \frac{1}{(2 \pi \Delta\tau^2)^{3/2}}
        \exp\left[-\frac{(\tau - \tau_0)^2} {2  \Delta \tau^2} \right],
$
where $\tau_0$ is the mean freeze-out time, and the $\Delta\tau$ is the
duration of particle emission, satisfying $\Delta\tau \ll \tau_0$.
The (inverse) Boltzmann phase-space distribution, $B(x,k)$ is 
given by 
\begin{equation}B(x,k)=
  \exp\left( \frac{ k^\nu u_\nu(x)}{T(x)} -\frac{\mu(x)}{T(x)} \right),
\end{equation}
and the term $s_q$ is 
$ 0$, $-1$, and $+1$  for Boltzmann, Bose-Einstein and Fermi-Dirac
statistics, respectively.
The flow four-velocity,
$u^\nu(x)$, the chemical potential, $\mu(x)$, and the temperature, $T(x)$
distributions for axially symmetric collisions were  
determined from the principles of simplicity, analyticity and 
correspondence to hydrodynamical solutions in the limits when
such solutions were known~\cite{Csorgo:1995bi,Csorgo:1995vf}. 
Recently, the Buda-Lund hydro model 
lead to the discovery of a number of new, exact analytic
solutions of hydrodynamics, 
both in the relativistic~\cite{relsol-cyl,relsol-ell}
and in the non-relativistic domain~\cite{nr-sol,nr-ell,nr-inf}.

The expanding matter is assumed to follow a three-dimensional,
 relativistic flow, characterized by
transverse and longitudinal Hubble constants,
\be
        u^{\nu}(x)  =  \left( \gamma, H_t r_x, H_t r_y, H_z r_z \right),
\ee
where $\gamma$ is given by the normalization
condition $u^\nu(x) u_\nu(x) = 1$. 
In the original form, this four-velocity distribution $u^\nu(x)$ 
was written as a linear transverse flow, superposed on
a scaling longitudinal Bjorken flow . The strength of the transverse flow
was characterized by its value $\langle u_t\rangle$ at the
``geometrical" radius $R_G$, 
see refs.~\cite{Csorgo:1995bi,Chapman:1994ax,Ster:1998hu}:
\bea
        u^{\nu}(x) & = & \left( \cosh[\eta] \cosh[\eta_t],
        \, \sinh[\eta_t]  \frac{r_x}{r_t},
        \, \sinh[\eta_t]  \frac{r_y}{r_t},
        \, \sinh[\eta] \cosh[\eta_t] \right), \nonumber \\
        \sinh[\eta_t]   & = & \ave{u_t} r_t / R_G,
\eea
with $ r_t = (r_x^2 + r_y^2)^{1/2}$.  Such a flow profile, with a
time-dependent radius parameter $R_G$, was recently shown to be an
exact solution of the equations of relativistic hydrodynamics of a
perfect fluid at a vanishing speed of sound,  see refs.~\cite{Biro:1999eh,Biro:2000nj}.

The Buda-Lund hydro model characterizes the inverse temperature 
$1/T(x)$, and fugacity, $\exp\left[\mu(x)/T(x)\right]$
distributions of an axially symmetric, finite
hydrodynamically expanding system with the mean and the variance of
these distributions, in particular 
\bea
        \frac{\mu(x)}{T(x)} & = & \frac{\mu_0}{T_0} -
        \frac{ r_x^2 + r_y^2}{2 R_G^2}
        -{ \dst (\eta - y_0)^2 \ov 2 \Delta \eta^2 }, \label{e:mu} \\
        {\dst 1 \ov T(x)} & =  &
        {\dst 1 \ov T_0 } \,\,
        \left( 1 + {\dst  r_t^2 \ov 2 R_s^2} \right) \,
        \left( 1 + {\dst (\tau - \tau_0)^2 \ov 2 \Delta\tau_s^2  } \right).
\eea
	Here $R_G$ and $\Delta\eta$ characterize the spatial scales
	of variation of the fugacity distribution, $\exp\left[\mu(x)/T(x)\right]$,
	that control particle densities. Hence these scales are
	referred to as geometrical lengths. These are distinguished from
	the scales on which the inverse temperature distribution changes,
	the temperature drops to half if $r_x = r_y = R_s$ or 
	if $\tau = \tau_0 + \sqrt{2} \Delta\tau_s$. These parameters
	can be considered as second order Taylor expansion coefficients
	of these profile functions, restricted only by the symmetry
	properties of the source, and can be trivially expressed 
	by re-scaling the earlier fit parameters.
	The above is the most direct form of the Buda-Lund model.
	However,  different combinations may also be used to measure
	the flow, temperature and fugacity profiles~\cite{Csorgo:1995bi,cs-rev}:
	 $ H_t   \equiv  {b}/{\tau_0} \, = \, \ave{u_t} / R_G
	\, = \, \ave{u_t^\prime} / R_s$ ,  $H_l   \equiv \gamma_t /\tau_0$,
	where $ \gamma_t = \sqrt{ 1  + H_t^2 r_t^2}$ is evaluated 
	at the point of maximal emittivity, and  
\bea
        \frac{1}{ R_s^2} & =  & \frac{a^2}{\tau_0^2} \, = \, 
		 \ave{\frac{\Delta T}{T}}_r  \frac{1}{R_G^2} \, = \, 
		 \frac{ T_0 - T_s}{T_s}  \frac{1}{R_G^2}, \\
        \frac{1}{ \Delta\tau_s^2} & = & \frac{d^2}{\tau_0^2} \, = \, 
		 \ave{\frac{\Delta T}{T}}_s  \frac{1}{\Delta\tau^2} \, = \, 
		 \frac{ T_0 - T_e}{T_e}  \frac{1}{\Delta\tau^2}. 
\eea

\section{Buda-Lund fits to Au+Au data at $\sqrt{s_{\NN}}= 200$ GeV }
       	In this section, we present new fit results to
	BRAHMS data on charged particle pseudorapidity
        distributions~\cite{Bearden:2001qq}, 
	and PHENIX data on identified particle momentum distributions
        and Bose-Einstein (HBT) radii~\cite{Adler:2003cb,Adler:2004rq}
        in Au+Au collisions at $\sqrt{s_{\NN}}= 200$ GeV.

	The analysis codes and methods are identical to the ones
        used to fit the BRAHMS~\cite{Bearden:2001xw},
        PHENIX~\cite{Adcox:2001mf,Adcox:2002uc},
        PHOBOS~\cite{Back:2001bq}, and STAR~\cite{Adler:2001zd}
         data in 0- 5\% most central Au+Au collisions
	at $\sqrt{s_{\NN}} = 130$ GeV, see ref.~\cite{ster-ismd03}.
	The applied Buda-Lund 1.5 fitting package
	can be downloaded, together  with the detailed fit results, from 
	ref.~\cite{Csorgo-blhome}.  
        This calculation determines
	the position of the saddle point exactly in the
        beam direction, but in the transverse direction, the saddle
        point equations are solved only approximately,
	as summarized in ref.~\cite{cs-rev}. 

	The new results for $\sqrt{s_{\NN}} = 200$ GeV  Au+Au  
	collisions in the 0-30\% centrality class
	are shown in the first column of  Table 1.
	For comparison, we also show the results 
	of an identical fit to $\sqrt{s_{\NN}} = 130$ GeV  Au+Au
        collisions in the 0- 5\% centrality class.

\begin{table}[b]
\begin{center}
\begin{tabular}{|l|rl|rl|rl|rl|}
\hline
\hline
                \null 
                 & \multicolumn{2}{c|}{Au+Au \@ 200 GeV}
                 & \multicolumn{2}{c|}{Au+Au \@ 130 GeV} \\ 
                Buda-Lund 
                 & \multicolumn{2}{c|}{BRAHMS+}
                 & \multicolumn{2}{c|}{BRAHMS+PHENIX} \\ 
                v1.5 
                 & \multicolumn{2}{c|}{PHENIX}
                 & \multicolumn{2}{c|}{+PHOBOS+STAR} \\ 
\null			& 0 - & 30 \%  \null
                      & 0 - & 5(6) \% \null
		\\
		\hline
		\hline
$T_0$ [MeV]           & \hspace{0.3cm} 200    &$\pm$ 9
                      & \hspace{0.3cm} 214    &$\pm$ 7
		\\
$T_{\mbox{\rm e}}$ [MeV]
                      & 127    &$\pm$ 13
                      & 102    &$\pm$ 11
		\\ 
$\mu_B$ [MeV]	
		      & 61	& $\pm$  40
		     & 77 	& $\pm$  38 
		\\ \hline
$R_{G}$ [fm]          & 13.2   &$\pm$ 1.3
                      & 28.0   &$\pm$ 5.5
		\\
$R_{s}$ [fm]          & 11.6    &$\pm$ 1.0
                      & 8.6    &$\pm$ 0.4
		\\
$\langle u_t^\prime \rangle$ 
	              & 1.5   &$\pm$  0.1
                      & 1.0   &$\pm$  0.1
		\\ \hline
$\tau_0$ [fm/c]       & 5.7    &$\pm$ 0.2
                      & 6.0    &$\pm$ 0.2 
		\\
$\Delta\tau$ [fm/c]   & 1.9    &$\pm$ 0.5
                      & 0.3    &$\pm$ 1.2
		\\
$\Delta\eta$          & 3.1    &$\pm$ 0.1
                      & 2.4    &$\pm$ 0.1
		\\ \hline
$\chi^2/\mbox{\rm NDF}$
                      & 132    &/ 208 
                      & 158.2  &/ 180 
		\\ \hline \hline
\end{tabular}
\end{center}
\caption
{
The first column shows the source parameters from simultaneous fits of
final BRAHMS and  PHENIX data for 0 - 30 \% most central 
$Au+Au$ collisions at $\sqrt{s_{\NN}} = 200$ GeV,
as shown in Figs. 1 and 2, 
as obtained  with the Buda-Lund hydro model, version 1.5.
The errors on these parameters are still preliminary.
The second column is the result of an identical analysis of
BRAHMS, PHENIX, PHOBOS and STAR data for 0 - 5 \% most central
Au+Au collisions at $\sqrt{s_{\NN}}=130$ GeV, ref.~\cite{ster-ismd03}.
}
\label{tab:results}
\end{table}

\begin{figure}[!thb]
\begin{center}
  \begin{center}
	\includegraphics[width=2.7in]{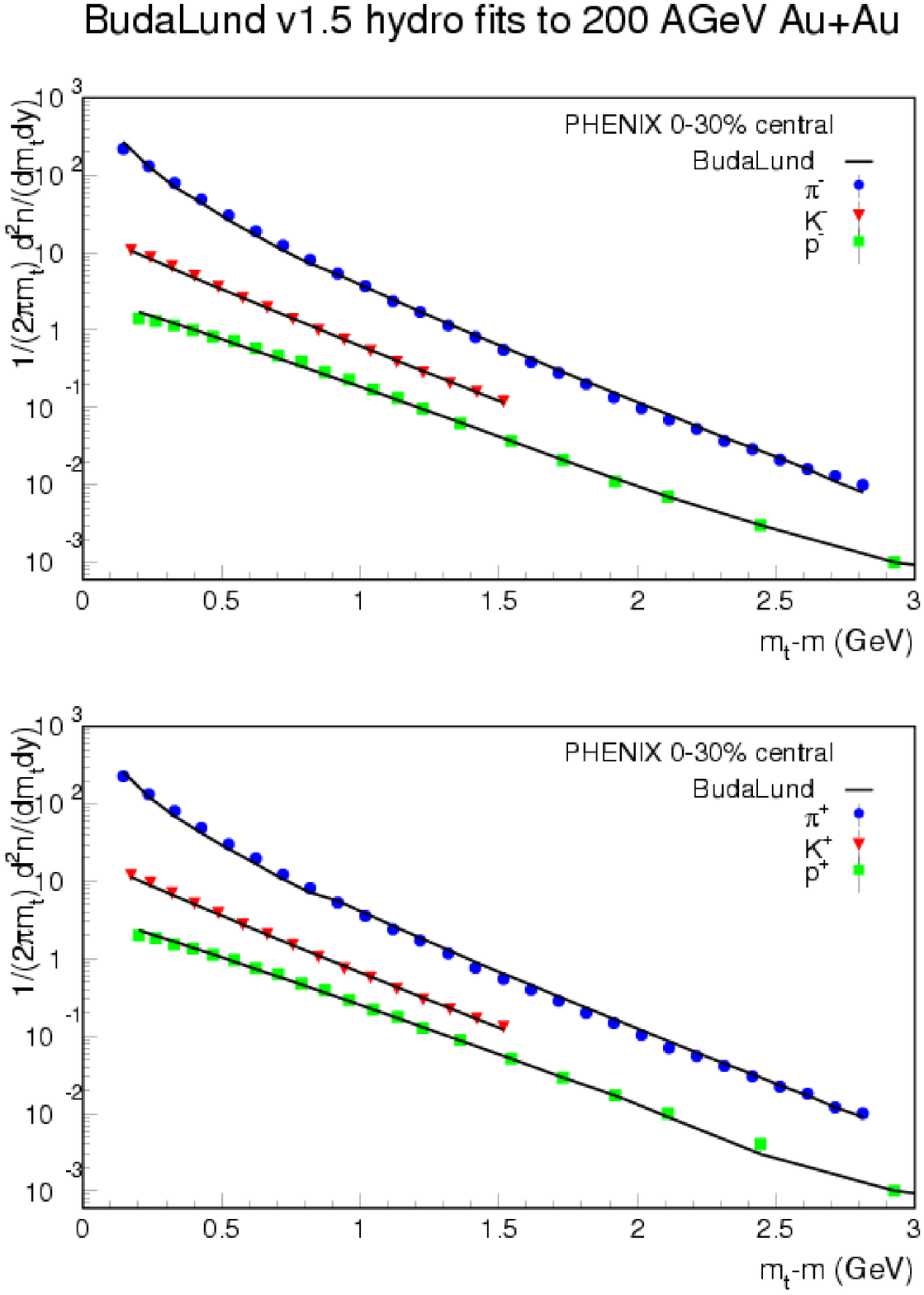}
	\includegraphics[width=2.7in]{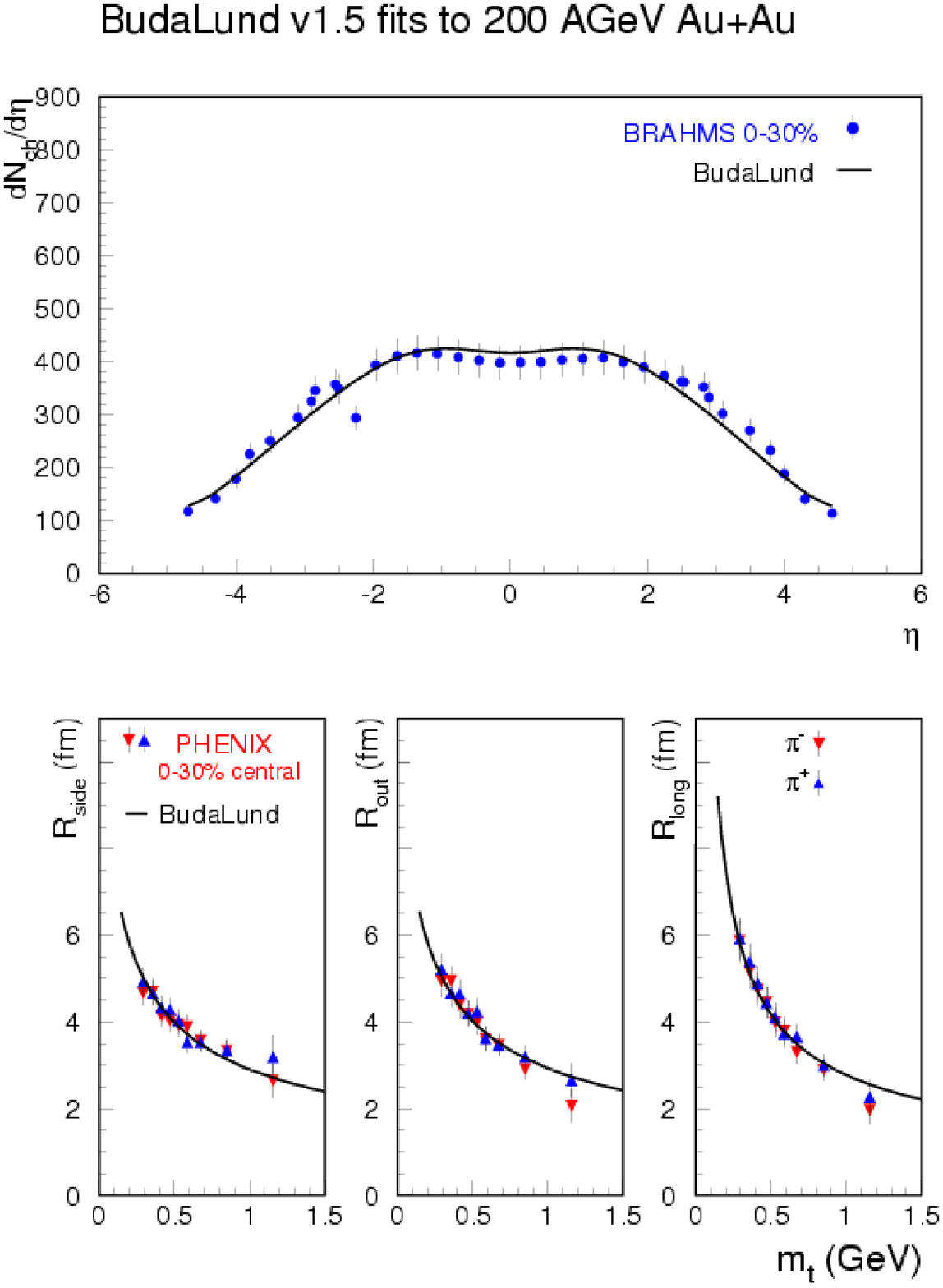}
\end{center}
\caption[*]{
\label{fig:spectra}
{\small Solid line shows the simultaneous Buda-Lund v1.5 fit  to final
Au+Au data at $\sqrt{s_{\NN}} = 200$ GeV. The 
transverse mass distributions of identified particles are 
measured by PHENIX~\cite{Adler:2003cb}
the  pseudorapidity distributions of charged particles
are measured by BRAHMS~\cite{Bearden:2001qq},
the transverse mass dependence of the radius parameters are
data of PHENIX~\cite{Adler:2004rq}. 
Note that the identified particle spectra are published
in more detailed centrality classes, but we recombined the 
0-30\% most central collisions so that the fitted spectra and 
radii be obtained in the same centrality class.} 
}
\end{center}
\end{figure}

\begin{figure}[!thb]
\begin{center}
  \begin{center}
\includegraphics[width=4.3in]{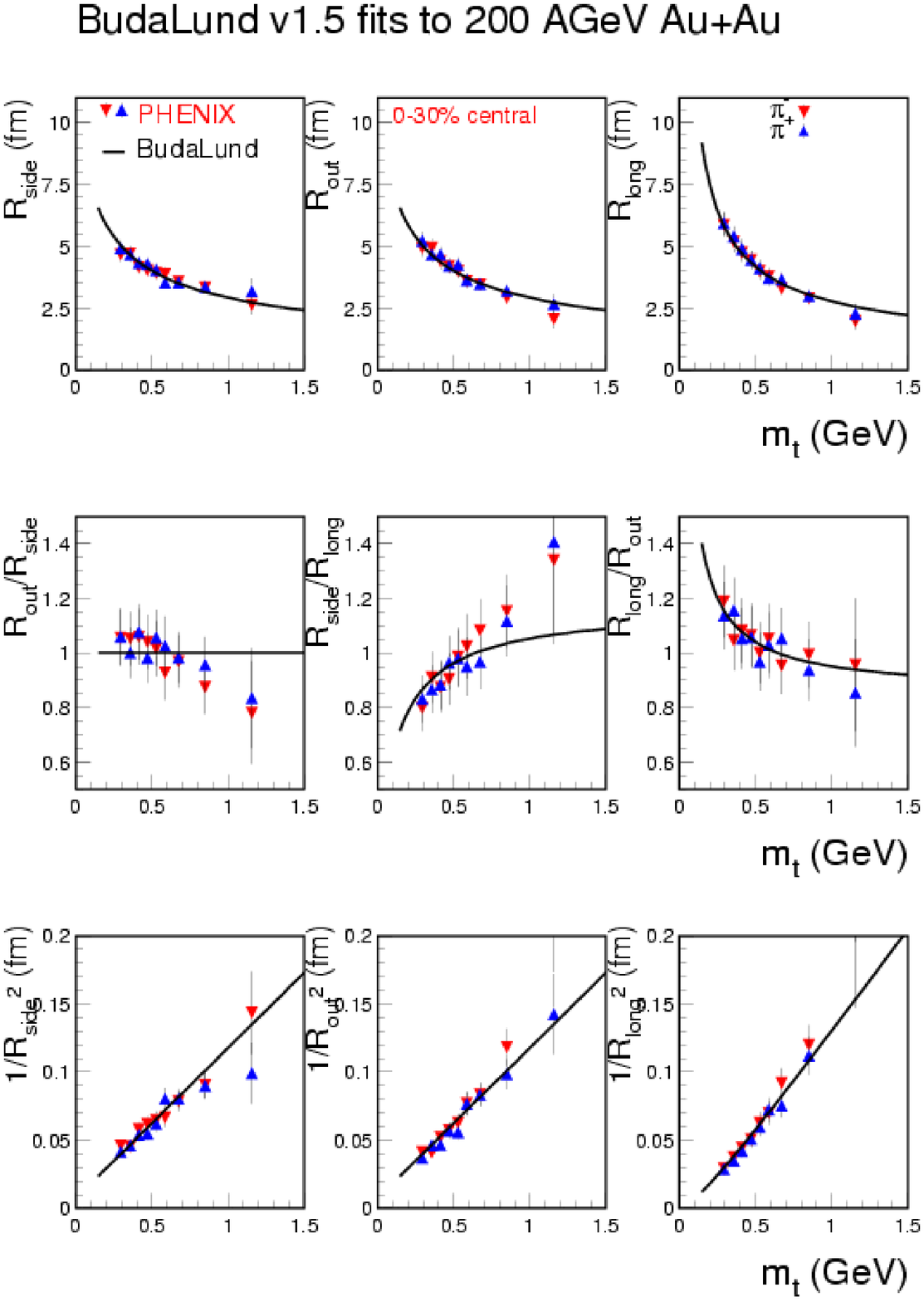}
\end{center}
\caption[*]{
\label{fig:radii} {\small Top row shows the transverse mass dependence of the
side, out and longitudinal HBT radii, the central line shows their
pairwise ratio (usually only $R_{\mbox{\rm out}}/R_{\mbox{\rm side}}$ is
shown) together with the  Buda-Lund fits, vers. 1.5. 
The bottom line shows the inverse of the squared radii.
The intercept of the curves in this row is within errors zero for the
two transverse components, so the fugacity is within errors 
independent of the transverse coordinates. 
However, the intercept is nonzero in the longitudinal direction, 
which makes the fugacity (hence particle ratios) rapidity dependent.
See also ref.~\cite{ster-ismd03} for a similar plot at $\sqrt{s_{\NN}} =
130$ GeV.}
}
\end{center}
\end{figure}

	Let us clarify first the meaning
	of the parameters shown in Table 1. 
	The temperature at the center of the fireball at 
	the mean freeze-out time is denoted by
	$T_0 \equiv T(r_x=r_y=0, \tau=\tau_0)$.
	The surface temperature is also a characteristic, kind of 
	an average temperature, and its value is always 
	$T_s \equiv T(r_x=r_y=R_s, \tau=\tau_0) = T_0/2$. 
	In fact this relationship defines the ``surface" radius $R_s$. 
	During the particle emission,
	the system may cool due to evaporation and expansion, this is
	measured by the ``post-evaporation temperature" 
	$T_e \equiv T(r_x=r_y=0, \tau = \tau_0 + \sqrt{2} \Delta\tau)$. 
 	In the presented cases, 
	the strength of the transverse flow is measured
	by $\ave{u_t^\prime}$, its value at the ``surface radius" $R_s$. 
	The ``mean freeze-out time" parameter
	is denoted by $\tau_0$ and  the ``duration" of particle
	emission, or the  width of the freeze-out time distribution
	is measured by $\Delta\tau$. 
	The fugacity distribution varies on the characteristic
	transverse scale given by the ``geometrical radius" $R_G$.
	Finally, the width of the space-time rapidity distribution, 
	or the longitudinal variation scale of the fugacity 
	distribution is measured by the parameter $\Delta\eta$.

	Perhaps it could be more appropriate
	to directly fit the transverse Hubble constant, $H_t = 
	\ave{u_t^\prime}/R_s$ to the data, as this value is 
	not sensitive to the length-scale chosen to evaluate 
	the ``average" transverse flow $\ave{u_t^\prime}$. 	
	In the case of parameters shown in Table 1, the density
	drop in the transverse direction is dominated by the
	cooling of the local temperature distribution in the transverse
	direction, and not so much by the change of the fugacity
	distribution. That is why we fitted here $\ave{u_t^\prime}$ at
	the ``surface radius" $R_s$.
	Note also that  $\tau_0$ could more properly be interpreted 
	as the inverse  of the  longitudinal Hubble constant $H_l$,
	 which is only an order of magnitude estimate of the mean 
	freeze-out time, similarly to how the inverse of the present
	 value of the Hubble constant in astrophysics provides 
	only an order of magnitude estimate of the life-time of our Universe.
	The feasibility of directly fitting the transverse and
	longitudinal Hubble constants to data will be investigated
	in a subsequent publication. 

	Let us also note, that we have fitted the absolute normalized
	spectra for identified particles, and the normalization conditions
	were given by central chemical potentials $\mu_0$ that were
	taken as free normalization parameters for each particle species.
	All these directly fitted parameters are 
	made public at~\cite{Csorgo-blhome}.
	From these values, we have determined the net bariochemical potential as
	$\mu_B = \mu_p - \mu_{\overline{p}}$. Although this parameter
	is not directly fitted but calculated,
	we have included $\mu_B$ in Table 1,
	so that our results could be compared with other successful 
	models of two-particle Bose-Einstein correlations
	at RHIC, namely the AMPT cascade~\cite{Lin:2002gc}, 
	Tom Humanic's cascade~\cite{Humanic:2002iw},
	the blast-wave model~\cite{Retiere:2003qb,Retiere:2003kf}, 
	the Hirano-Tsuda numerical hydro~\cite{Hirano:2002hv}
	and the  Cracow ``single freeze-out thermal model"
	~\cite{Broniowski:2002wp,Florkowski:2002wn,Broniowski:2001we}.

	Now, we are ready for the discussion of the results in Table 1.
	In case of more central collisions at the  lower RHIC energies, a
well defined minimum was found, with accurate error matrix and a
statistically acceptable fit quality, $\chi^2$/NDF=
158/180, that corresponds to a confidence level of 88 \%. 
(These fit results were shown graphically
on Figs. 1  and 2 of ref.~\cite{ster-ismd03}, and the parameters
are summarized in the second column of Table 1.) 
In the case of the less central but more energetic Au+Au collisions,
the obtained $\chi^2/\mbox{\rm NDF}$ fit is {\it too small}.
Note that in these fits we added the systematic and 
statistical errors in quadrature, and this procedure is preliminary and
 has to be revisited before we can report on the final values of the fit
parameters and determine their error bars.
It could also be advantageous to analyze a more central 
data sample, or the centrality dependence of the radius
parameters and the pseudorapidity distributions,
or to fit additional data of STAR and PHOBOS too,
so that the parameters of the Buda-Lund hydro model
 could be determined with smaller error bars.

At present, we find that $T_0 > T_c = 172 \pm 3$ MeV~\cite{Fodor:2001pe} 
by 3 $\sigma$ in case of the 0-30 \% most central Au+Au 
data at $\sqrt{s_{\NN}} = 200$ GeV,
while $T_0 > T_c$ by more than 5 $\sigma$ in
case of the 0-5(6) \% most central Au+Au data at $\sqrt{s_{\NN}} = 130 $ GeV. 
Thus this signal of a cross-over transition to 
quark deconfinement is not yet significant in the more energetic 
but less central Au+Au data sample, while it is significant
at the more central, but less energetic sample.
In this latter case of 130 GeV Au+Au data,
$R_G$ obviously became an irrelevant parameter, with $1/R_G\approx 0.$ .
This is explicitly visible in Fig. 2 of ref.~\cite{ster-ismd03},
where the last row indicates that the correlation radii are 
in the scaling limit and the fugacity distribution,
$\exp\left[\mu(x)/T(x)\right]$ is independent of the
transverse coordinates. 

The Buda-Lund model predicted, see eqs. (53-58) in 
ref.~\cite{Csorgo:1995bi} and also eqs. (26-28) in~\cite{nr-inf}, 
that the linearity of the inverse radii as a function of $m_t$ 
can be connected to the Hubble flow and the temperature gradients. 
The slopes are the same for side,
out and longitudinal radii if the Hubble flow (and the temperature
inhomogeneities) become direction independent. The intercepts of the
linearly extrapolated $m_t$ dependent inverse squared radii at $m_t=0$ determine
$1/R_G^2$, or  the magnitude of corrections from the finite
geometrical source sizes, that stem from the $\exp[\mu(x)/T(x)]$
terms. We can see on Fig. 2, that these corrections within errors
vanish also in $\sqrt{s_{\NN}} = 200$ Au+Au collisions
at RHIC. This result is important, because it explains, why
thermal and statistical models are successful at RHIC: if
 $\exp[\mu(x)/T(x)] = \exp(\mu_0/T_0)$, then this factor becomes
an overall normalization factor, proportional to the particle
abundances. Indeed, we found that when the finite size in the
transverse direction is generated by the $T(x)$ distribution,
the quality of the fit increased and we had no degenerate
parameters in the fit any more. 
This is also the reason, why we interpret $R_s$,
given by the condition that $T(r_x=r_y=R_s) = T_0/2$,
as a ``surface" radius: this is the scale where particle density drops.

Note that we have obtained similarly good description of
these data if we require that the four-velocity field is a fully
developed, three-dimensional Hubble flow, with $u^\nu =
x^\nu/\tau$, however, we cannot elaborate on this point here due
to the space limitations~\cite{mate-ell1}.

\section{Conclusions}
Table 1, Figures 1 and 2 indicate 
that the Buda-Lund hydro model works well
both at the lower and the higher RHIC energies,
 and gives a good quality description of the
transverse mass dependence of the HBT radii. For the dynamical reason,
see refs.~\cite{nr-inf} and \cite{Csorgo:1995bi}. In fact, even the time
evolution of the entrophy density can be solved from the fit results,
$s(\tau) = s_0 (\tau_0/\tau)^3$, which is the consequence 
of the Hubble flow, $u^\nu = x^\nu/\tau$, a well known solution of
relativistic hydrodynamics, see also ref.~\cite{relsol-ell}. 
This is can be considered as the resolution of the RHIC HBT ``puzzle",
although a careful search of the literature indicates that this
``puzzle" was only present in models that were not tuned to CERN SPS 
data~\cite{Csorgo:2002ry}.

We also observe that the central temperature is 
 $T_0 = 214 \pm 7$ MeV in the most central
Au+Au collisions at $\sqrt{s_{\NN}} = 130$ GeV, and we find here
a net bariochemical potential of $\mu_B = 77 \pm 38$ MeV. 
Recent lattice QCD results indicate~\cite{Fodor:2001pe}, that 
the critical temperature is within errors a constant of 
 $T_c = 172 \pm 3$ MeV in the $0 \le \mu_B \le 300$ MeV interval. 
 Our results clearly indicate $(T,\mu_B)$ 
values above this critical line, which is 
a significant, more than 5 $\sigma$ effect.
The present level of precision and the currently fitted PHENIX
and BRAHMS data set does not yet allow a firm conclusion about
such an effect at $\sqrt{s_{\NN}} = 200$ GeV, however,
a similar behavior is seen on a 3 standard deviation level.
This can be interpreted as a hint at quark deconfinement
at $\sqrt{s_{\NN}} = 200$ GeV at RHIC.

Finding similar parameters from the analysis of the pseudorapidity
dependence of the elliptic flow, it was estimated in
ref.~\cite{mate-ell1} that 1/8th of the total volume is above the
critical temperature in Au+Au collisions at $\sqrt{s_{\NN}} = 130$ 
GeV, at the time when pions are emitted from the source. 
We interpret this result as an
indication for quark deconfinement and a cross-over transition in
Au+Au collisions at $\sqrt{s_{\NN}} = 130 $ GeV at RHIC. 
This result was signaled first in ref.~\cite{Csorgo:2002ry}
in a Buda-Lund analysis of the final PHENIX and STAR data on midrapidity
spectra and Bose-Einstein correlations, but only at a three
standard deviation level. By including the pseudorapidity
distributions of BRAHMS and PHOBOS, the $T_0 \gg T_c$ effect
became significant in most central 
Au+Au collisions at $\sqrt{s_{\NN}} = 130 $ GeV. 
We are looking forward to observe, what happens with the present  signal 
in Au+Au collisions at $\sqrt{s_{\NN}} = 200$ GeV,
if we include STAR and PHOBOS data to the fitted sample.

The above observation of temperatures,
that are higher than the critical one, 
is only an indication, with other words, an indirect proof
for the production of a new phase, 
as the critical temperature is not extracted directly from the data,
but taken from recent lattice QCD calculations.

More data are needed to clarify the picture of quark deconfinement 
at the maximal RHIC energies, for example the centrality dependence
of the Bose-Einstein (HBT) radius parameters could provide very
important insights.

\section*{Acknowledgments}
T. Cs. and M. Cs. would like to the Organizers for their kind hospitality
and for their creating an
inspiring and fruitful meeting.
The support of the following grants are gratefully acknowledged:
OTKA T034269, T038406, the OTKA-MTA-NSF grant INT0089462, the
NATO PST.CLG.980086 grant and the exchange program 
of the Hungarian and Polish Academy of Sciences.

\end{document}